\documentclass[11pt,a4paper]{article}
\usepackage{amsmath,amsfonts,amssymb}
\usepackage{rotating,graphicx}
\usepackage{epic,eepic}

\numberwithin{equation}{section}

\renewcommand{\title}[1]{{\Large\bf\mbox{}\\\medskip#1\bigskip\medskip\\}}
\renewcommand{\author}[1]{{\large #1\smallskip\\}}
\newcommand{\address}[1]{{\em #1\medskip\\}}
\newcommand{\sm}[1]{{\scriptstyle #1}}
\newcommand{\smb}[2]{\makebox(0,0)[#1]{$\sm{#2}$}}
\newcommand{\W}[5]{W\!\left(\,\begin{array}{@{}cc|@{\:}}#4&#3\\
  #1&#2\end{array}\;#5\right)}
\newcommand{\face}[5]{\begin{picture}(1.6,1.6)
  \multiput(0.3,0.3)(1,0){2}{\line(0,1){1}}
  \multiput(0.3,0.3)(0,1){2}{\line(1,0){1}}
  \put(0.26,0.26){\smb{tr}{#1}}\put(1.34,0.26){\smb{tl}{#2}}
  \put(1.34,1.34){\smb{bl}{#3}}
  \put(0.26,1.34){\smb{br}{#4}}
  \put(0.8,0.8){\smb{}{#5}}\end{picture}}
\newcommand{\B}[6]{B^{#1 #2}\!\left(#3\;\;\;\begin{array}{@{}c|@{\:}}#4\\
  #4\end{array}\;#5,\,#6\right)}
\newcommand{\boundary}[3]{
  \begin{picture}(2.6,2.6)(-0.2,0.1)
  \put(0.5,1.5){\dashline[18]{0.2}(0,0)(1,1)(1.5,1)(1.5,-1)(1,-1)(0,0)}
  \put(0.34,1.5){\makebox(0,0)[r]{$\scriptstyle #1$}}
  \put(1.4,2.7){\makebox(0,0){$\scriptstyle #2$}}
  \put(1.4,0.3){\makebox(0,0){$\scriptstyle #2$}}
  \put(1.4,1.5){\makebox(0,0){#3}}
  \end{picture}}

\newcommand{\re}{\mbox{Re}}
\newcommand{\im}{\mbox{Im}}

\newcommand{\be}{\begin{equation}}
\newcommand{\ee}{\end{equation}}
\newcommand{\bea}{\begin{eqnarray}}
\newcommand{\eea}{\end{eqnarray}}
\newcommand{\I}{\boldsymbol{I}}
\renewcommand{\v}{|0\rangle}
\newcommand{\gs}{|\frac{1}{10}\rangle}
\newcommand{\xilatt}{\xi_{\mbox{\scriptsize latt}}}
\makeatletter
\renewcommand{\@makecaption}[2]{
   \vskip\abovecaptionskip
   \sbox\@tempboxa{#1. #2}%
   \ifdim \wd\@tempboxa >\hsize
     #1. #2\par
   \else
     \global \@minipagefalse
     \hb@xt@\hsize{\hfil\box\@tempboxa\hfil}%
   \fi
   \vskip\belowcaptionskip}
\makeatother

\begin{document}

\vspace*{-15mm}
\begin{flushright} SISSA 33/2004/FM
\end{flushright}
\begin{center}

\title{Integrals of motion from TBA  
and lattice-conformal dictionary}
  
\renewcommand{\thefootnote}{\fnsymbol{footnote}}
\author{Giovanni Feverati\footnote[1]{feverati@sissa.it},
Paolo Grinza\footnote[7]{grinza@sissa.it}}

\address{International School for Advanced Studies (SISSA)\\
via Beirut 2-4, 34014 Trieste, Italy\\
INFN, sezione di Trieste}

\end{center}
\renewcommand{\thefootnote}{\arabic{footnote}}
\setcounter{footnote}{0}

\begin{abstract}
The integrals of motion of the tricritical Ising model 
are obtained by Thermodynamic Bethe Ansatz (TBA) equations derived from 
the $A_4$ integrable lattice model. They are compared with those given 
by the conformal field theory leading to a unique one-to-one 
lattice-conformal correspondence.  
They can also be followed along the renormalization group flows generated 
by the action of the boundary field $\varphi_{1,3}$ on
conformal boundary conditions in close analogy to the usual TBA description of 
energies.
\end{abstract}

\section{Introduction}
The study of two-dimensional integrable quantum field theories (QFTs) 
and integrable lattice 
models has lead to the discovery of a large number of deep analogies between 
them. It is worth to recall here that, in both cases, integrability is 
equivalent to an 
algebraic relation (Yang-Baxter equation) for the fundamental ``evolution''
operator in the two languages, the S-matrix in the former case
and the transfer matrix in the latter case. 
Furthermore, a possible link between them is given by the so-called
conformally invariant field 
theories (CFTs) because they play a fundamental unifying role as being 
themselves integrable and describing the critical properties of lattice models.
Hence, adopting this point of view, one can consider integrable QFTs obtained 
by adding a perturbation which preserves the integrable structure of the conformal
field theory. 

The relevant case, here, is the $\varphi_{1,3}$ boundary perturbation 
of the $c=7/10$ unitary minimal model (it describes the universality class 
of the tricritical Ising model (TIM)).
Indeed, if this theory is defined on a cylinder of finite length, the operator 
$\varphi_{1,3}$ acting on the boundaries generates a set of integrable flows
\cite{affleck} between different sectors of the model. 

On the other hand, many results indicate that the $A_4$ Andrews-Baxter-Forrester 
(ABF) integrable lattice model \cite{ABF} belongs to the same universality class 
of TIM. In particular, the boundary flows of TIM were extensively studied in 
\cite{FPR2, F} using the continuum scaling limit of this model
(see \cite{LeClair:1995uf,Dorey:1997yg} and references in \cite{FPR2} for the derivation of the TBA equations). 
Moreover, this 
correspondence has been made stronger by showing that the scaling limit of both
the energy and the associated degeneracies of the lattice model exactly matches 
with the corresponding quantities given by the Virasoro characters of the 
$c=7/10$ minimal unitary CFT \cite{OPW}. 

The aim of this paper is to establish a \emph{lattice-conformal dictionary}
(the wording is borrowed from \cite{melzer}) 
between the lattice model and the (possibly perturbed) underlying CFT. 
In other words we will 
find a correspondence between the states on the lattice, given by the 
eigenvalues/eigenvectors of the transfer matrix, and the states which belong to the Virasoro
irreducible modules. In order to achieve this result we will look at a suitable
family of involutive local integrals of motion (IM). From the point of view of
CFT such integrals of motion can be constructed as appropriate 
polynomials of the 
Virasoro generators whose actual expression can be obtained in a constructive way
(since a general expression is not known to exist), see \cite{SY}. 
Within the lattice model, such 
integrals of motion can be accessed by usual TBA techniques showing a remarkable
correspondence between their eigenstates and those of the transfer matrix (a similar analysis was done in \cite{Fioravanti:2002sq} by means of the nonlinear integral equations method in the context of KdV theory.).
Therefore, as we will show in the sequel, the desired state-by-state correspondence 
can be found by comparing the eigenvalues of the integrals of
motion in the CFT with those of the lattice model. 
In this sense, our approach is algebraic-analytic while in 
\cite{FinChar} it is combinatorial. 

Finally, one can wonder about the fate of such integrals of motion when a 
relevant perturbation is switched on. In the present case, where the 
$\varphi_{1,3}$
boundary perturbation is concerned, the answer is that they 
can be chosen to be preserved, i.e. 
to remain involutive integrals of motion under such a perturbation. 
Indeed, among the possible families of involutive integrals of 
motion allowed in the CFT, we will consider that one whose spins are 
predicted to be preserved, for the present perturbation,
by the counting argument in \cite{zam}. 
This implies that the perturbed theory is still integrable
and it is remarkable that the eigenvalues of the integrals of motion
can be followed along the Renormalization Group flow in close 
analogy to the usual TBA description of energies.

As a final consideration, we also find a connection between our results and a 
recent series of papers by Bazhanov, Lukyanov and Zamolodchikov (BLZ)
\cite{blz94,blzxx}. We will see that, up to normalizations, the eigenvalues 
of the transfer matrix (and hence those of the integrals of motion) of the 
lattice model coincides with the eigenvalues of the so-called 
``{\bf T}-operators'' introduced in the above mentioned papers.   

The outline of the paper is as follows. In sect.~2 we describe the $A_4$ 
ABF integrable lattice model; in sect.~3 we 
carefully discuss both the continuum scaling limit of the lattice 
transfer matrix and 
the emergence of the integrals of motion in such a context;
sect.~4 is devoted to a brief presentation of the integrals of motion in 
CFT. We give our results in sect.~5. 
The connection between our paper and the BLZ papers is elucidated in
sect.~6, and finally we draw our conclusions in sect.~7.

\section{$A_4$ lattice model}
This model was exhaustively discussed in \cite{OPW, FPR2} and we 
summarize here the properties that will be useful later.  
For convenience, we will adopt the notations introduced in \cite{FPR2}.
At the critical temperature, on a cylinder of finite 
length, it is defined by the following double row transfer matrix 
\setlength{\unitlength}{10mm}
\begin{eqnarray}
& &\mathbf{D}^{N}(u,\xi)_{\sigma,\sigma'}\equiv
\mathbf{D}^{N}_{1,1|2,1}(u,\xilatt)_{\sigma,\sigma'}
=\displaystyle\sum_{\tau_{1},\dots,\tau_{N}}
\raisebox{-1.4\unitlength}[1.3\unitlength][1.1\unitlength]
{\begin{picture}(7,2.2)(0.7,0.05)
\put(6,1.5){\dashline[18]{0.2}(0,0)(1,1)(1.5,1)(1.5,-1)(1,-1)(0,0)}
\put(6,0.5){\dottedline[\circle*{0.013}]{0.18}(0,0)(1,0)}
\put(6,2.5){\dottedline[\circle*{0.013}]{0.18}(0,0)(1,0)}
\multiput(1,0.5)(1,0){3}{\line(0,1){2}}
\multiput(5,0.5)(1,0){2}{\line(0,1){2}}
\multiput(1,0.5)(0,1){3}{\line(1,0){5}}
\put(1,0.35){\smb{t}{1}}
\put(2,0.35){\smb{t}{\sigma_1}}\put(3,0.35){\smb{t}{\sigma_2}}
\put(5,0.35){\smb{t}{\sigma_{N-1}}}\put(6,0.35){\smb{t}{2}}
\put(7,0.35){\smb{t}{2}}
\put(1,2.6){\smb{b}{1}}
\put(2,2.6){\smb{b}{\sigma'_{1}}}\put(3,2.6){\smb{b}{\sigma'_2}}
\put(5,2.6){\smb{b}{\sigma'_{N-1}}}\put(6,2.6){\smb{b}{2}}
\put(7,2.6){\smb{b}{2}}
\put(0.8,1.6){\smb{tl}{2}}\put(1.99,1.45){\smb{tr}{\tau_1}}
\put(2.99,1.45){\smb{tr}{\tau_2}}\put(4.99,1.45){\smb{tr}{\tau_{N-1}}}
\put(5.99,1.45){\smb{tr}{\tau_{N}}}
\multiput(1.5,1)(1,0){2}{\smb{}{u}}\put(5.5,1){\smb{}{u}}
\multiput(1.5,2)(1,0){2}{\smb{}{\lambda\!-\!u}}
\put(5.5,2){\smb{}{\lambda\!-\!u}}
\put(6.9,1.5){\makebox(0,0){\scriptsize $ u ,\,\xilatt $}}
\end{picture}} \nonumber \\[1mm]
& &\xilatt=-\lambda+\frac{i}{5}\xi, \qquad \xi \in \mathbb{R} \label{DRTMdef}
\end{eqnarray}
which has been specialized to the boundary flow that we are going to examine: 
$\chi_{1,2}\mapsto \chi_{1,1}$ (note that it is a constant $r$ flow).

The left boundary is kept fixed and the boundary interaction, shown with 
dashed lines, acts on the right boundary only and it is tuned by the 
parameter $\xi$. 
The heights $\sigma_j,\sigma'_j,\tau_j\in\{1,2,3,4\}$ on adjacent vertexes 
are constrained to follow the $A_4$ adjacency rule which also 
forces the number $N$ of faces 
on a row to be odd. In the following we will always 
restrict to this case, in particular when the limit 
$\lim _{N\rightarrow \infty}$ will be considered.
The bulk weights are fixed to their critical values
\begin{equation}\label{bulkw}
\W{a}{b}{c}{d}{u}=
\raisebox{-0.7\unitlength}[0.8\unitlength][0.7\unitlength]
{\ \face{a}{b}{c}{d}{u}}
\ = \frac{\sin(\lambda-u)}{\sin\lambda}\,\delta_{a,c}+
\frac{\sin u}{\sin\lambda}\sqrt{\frac{S_aS_c}{S_bS_d}}\,\delta_{b,d}
\end{equation}
where the physical range of the spectral parameter is $0<u<\lambda$, and 
$\lambda=\frac{\pi}{5}$ is the crossing parameter. The crossing factors
are defined as $S_a=\sin a\lambda / \sin \lambda$. \\
Among the general expressions for integrable boundaries 
given in \cite{BP}, we only need to consider
the following case:
\begin{eqnarray} 
\B{2}{,1}{2\pm 1}{2}{u}{\xilatt}&=&\raisebox{-1.3\unitlength}[1.3\unitlength]
{\boundary{2\pm 1}{2}{\scriptsize $ u, \; \xilatt $}}
=\sqrt{\frac{S_{2\pm 1}}{S_2}} \;
\frac{ \sin(\xilatt\pm u) \, \sin(2\lambda +\xilatt \mp u) }
{\displaystyle \sin \! \lambda \, \cosh2\im(\xilatt)} \nonumber \\ & = &
\raisebox{-1.3\unitlength}{\hspace{3mm}\begin{picture}(2.2,2.7)(0,0.1)
\multiput(0.5,0.5)(1,0){2}{\line(0,1){2}}
\multiput(0.5,0.5)(0,1){3}{\line(1,0){1}}
\put(0.5,0.3){\smb{}{2}} \put(0.1,1.5){\smb{}{2 \pm 1}}
\put(1.5,0.3){\smb{}{1}}
\put(1.5,2.7){\smb{}{1}}
\put(0.5,2.7){\smb{}{2}}
\put(1.03,1){\makebox(0,0){\scriptsize$\begin{array}{l} u-\lambda \\
-\xilatt \end{array}$}}
\put(1.03,2){\makebox(0,0){\scriptsize$\begin{array}{l} -u \\ -\xilatt
\end{array}$}}
\put(1.7,1.5){\smb{}{2}}
\end{picture}}
\sqrt{2\cos \lambda}\;\frac{-\sin\lambda}{\cos 2 \im (\xilatt)}.
\label{bweight_1}
\end{eqnarray}
The second line of the equation shows that the same boundary interaction 
can be obtained by adding a single column to the right of the lattice; 
the faces of this column depend upon $\xi$ but the new right boundary
is fixed to the sequence $1,2,1,\ldots$.

Let us now consider the right boundary of the lattice defined in (\ref{DRTMdef}).
We can see that, for $\xi=-\infty$ it is formed by the quasi-free sequence 
$2,2\pm 1,2,2\pm 1,\ldots$,
while for $\xi=0$ it is formed by the fixed sequence $1,2,1,2,\ldots$.
In the continuum scaling limit, the former case corresponds to 
the ultraviolet (UV) critical point $\chi_{1,2}(q)$ 
and the latter to the infrared (IR) critical point $\chi_{1,1}(q)$
of the boundary renormalization group flow. \\
By construction, the given bulk and boundary weights
satisfy the Yang-Baxter and Boundary Yang-Baxter equations leading
to an integrable model. Consequently, the double row transfer matrix forms 
a one parameter family of commuting operators:
\begin{equation}
[\mathbf{D}^{N}(u,\xi),\mathbf{D}^{N}(u',\xi)]=0
\end{equation}
for arbitrary complex values of the spectral parameters $u, u'$.
An important symmetry becomes apparent
when one defines the following normalized 
transfer matrix
\begin{eqnarray}
\label{normal}
\mathbf{t}^{N}(u,\xi) & = &
\mathbf{D}^{N}(u,\xi) \, S_{2}(u,\xilatt) \, S(u)\, \sqrt{2\cos \lambda}
\left[ \frac{\sin (u+2\lambda )\sin \lambda }
{\sin (u+3\lambda )\sin (u+\lambda )}\right] ^{2N}
\end{eqnarray}
where
\begin{eqnarray}
S(u)&=&\frac{\sin ^{2}(2u-\lambda )}{\sin (2u+\lambda )\sin (2u-3\lambda)},
\label{normal_s} \\[2mm]
S_{2}(u,\xilatt) & = &
\frac{-\sin \lambda \, \sin (u-\xilatt+\lambda )\sin (u+\xilatt+3\lambda )\,
\cosh2\im(\xilatt)}
{\sin (u-\xilatt)\sin (u-\xilatt+2\lambda )\sin (u+\xilatt-\lambda )
\sin(u+\xilatt+2\lambda)}.
\label{normal_h}
\end{eqnarray}
Indeed, one can show that the following functional equation is satisfied:
\begin{equation} \label{functional}
\mathbf{t}^{N}(u,\xi) \: \mathbf{t}^{N}(u+\lambda,\xi)=
1+ \mathbf{t}^{N}(u+3\lambda,\xi).
\end{equation}
The normalized transfer matrix $\mathbf{t}$ also forms a family of
commuting operators
\be \label{commt}
[\mathbf{t}^N(u,\xi),\mathbf{t}^N(u',\xi)]=0
\ee
so that the eigenstates are independent of $u$, and 
the functional equation holds true also for the eigenvalues.
We would like to stress that it is a crucial point since the TBA approach is 
precisely based on such an evidence.

In the continuum scaling limit, the large $N$ corrections to the
eigenvalues\footnote{We use $D,~t$ to indicate eigenvalues of
$\mathbf{D}, ~\mathbf{t}$.} of
the double row transfer matrix are related to the excitation energies of
the associated perturbed conformal field theory by
\begin{eqnarray} \label{leading}
-\frac{1}{2} \log D^{N}(u,\xi-\log N) &=&
N f_b(u) +f_{bd}(u,\xi)+\frac{2\pi \sin 5 u}{N}
\: E(\xi) +o(\frac{1}{N})
\end{eqnarray}
where $f_b$ is the bulk free energy, $f_{bd}$ is the surface
(i.e. boundary dependent) free energy and
$E(\xi)$ is a scaling function given by (\ref{scalingenergy}). 
Since both the bulk and surface 
free energy contributions
are the same for all the eigenvalues, they can be removed from the previous 
expression by subtracting the largest eigenvalue of $D^{N}$ (it plays
the role of a hamiltonian ground state). At the boundary critical points, 
the scaling energy $E(\xi)$ reduces to
\begin{equation} \label{crit}
E(\xi)=\left\{ \begin{array}{l@{\hspace{5mm}}l} 
-\frac{c}{24}+n,  & \text{IR:}~~\xi\rightarrow +\infty \\[3mm]
-\frac{c}{24}+\frac{1}{10}+n',  & \text{UV:}~~\xi\rightarrow -\infty 
\end{array} \right.
\end{equation}
where $n, n' \in \mathbb{N}$ are certain excitation levels. 

\begin{figure}\begin{center}
\setlength{\unitlength}{1mm}
\begin{picture}(70,46)
\put(0,0){\includegraphics[scale=0.5]{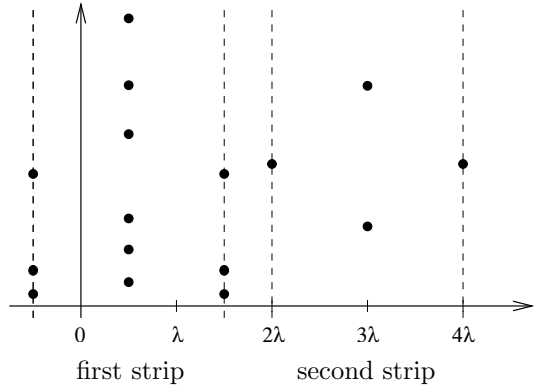}}
\put(9.2,-5){\small first strip}\put(38.5,-5){\small second strip}
\end{picture}\end{center}
\caption{Schematic example of zeros of the eigenvalue of the transfer 
matrix labeled $(2\,1\,1\,0\,0\,0|1\,0)$. 
Here: $m_1=6,\quad m_2=2,\quad n_1=3, \quad n_2=1 $.\label{figzeros}}
\end{figure}

The derivation of both the TBA equations and the scaling energy (\ref{crit})
is obtained by looking at the analytic structure of the eigenvalues
of the transfer matrix: $D^N(u,\xi)$ is an entire function of $u$ characterized
by simple zeros only, moreover it is periodic on the real axis $u\equiv u+\pi$. 
In the large $N$ limit, the zeros   
are organized in two sort of strings and they are distributed in two different strips, 
defined by $-\frac{1}{2}\lambda \leqslant \re(u)\leqslant \frac{3}{2}\lambda$ and 
$2\lambda \leqslant \re(u)\leqslant 4\lambda$ respectively. An example of such a 
situation is schematically shown in Fig.~\ref{figzeros}. \\
The zeros are bound either to appear in the center of each strip (located at 
$\re(u)=\frac{\lambda}{2} \text{~or~} 3\lambda$) and we will refer to them as 
\emph{1-strings},
or they can appear in pairs $(u,u')$ with the same imaginary part 
$\im(u)=\im(u')$ on the two edges of a strip, and we will call them \emph{2-strings}.
In order to label the number of such strings in the upper-half plane, 
we will use $m_1$ and $m_2$ for 1-strings located
in the first and second strip respectively; analogously $n_1$ and $n_2$ will be
used for 2-strings\footnote{We 
do not need to keep track of the lower half plane because it is the mirror
image of the upper half plane, thanks to the complex conjugation symmetry 
of the transfer matrix.}.
These numbers form the so-called 
$(\boldsymbol{m},\boldsymbol{n})$-system and are constrained to be
\begin{equation}\label{n1n2} \left.
\begin{array}{ccl}
n_1 &=& \frac{N+m_2}{2}-m_1 \geqslant 0 \\[2mm]
n_2 &=& \frac{m_1}{2}-m_2 \geqslant 0
\end{array} \right\} \Rightarrow m_1,m_2 ~~\mbox{even.}
\end{equation}
The imaginary part of the position of the 1-string is indicated by 
$v_k^{(j)}$ 
\be \text{first strip:} \quad
D^N(\frac{1}{2}\lambda+i v_k^{(1)},\xi)=0, \qquad \text{second strip:}\quad
D^N(3\lambda+i v_k^{(2)},\xi)=0;
\ee
where the $k$'s are ordered from the bottom to the top in the upper 
half plane, 
$0<v_1^{(j)}<v_2^{(j)}<\ldots<v_{m_1}^{(j)}$. Analogously, $w_k^{(j)}$ 
is used for the 2-strings.

Finally, the states (eigenvalues-eigenstates) both on the lattice and in the 
continuum scaling limit are uniquely characterized by the 
{\em non-negative quantum numbers} $I_k^{(j)}$. Their topological meaning 
is that when we refer to the $k$-th 1-string in strip $j$, 
$I_k^{(j)}$ is the number of 2-strings with larger imaginary part.
The notation which indicates the content of zeros works as in the following 
example as shown in Fig.~\ref{figzeros}: $(2\,1\,1\,0\,0\,0|1\,0)$ is a state with 6 zeros in the first
strip and 2 in the second; their quantum
numbers are respectively $I_1^{(1)}=2, \quad I_2^{(1)}=1, \ldots$.
In the sector $(1,2)$ a frozen zero can occur \cite{OPW}, so a parity 
$\sigma=\pm 1$ is added to the previous notation 
(namely $(\,\cdot\,|\,\cdot\,)_{\pm}$, see Tables 4 and 5).

\section{Continuous transfer matrix from the lattice}
Since our task is to compare lattice computations with CFT results, we have to define 
a suitable continuum limit of lattice quantities. 
In the following we will first define the continuum scaling limit for the 
lattice transfer 
matrix, and then the corresponding integrals of motion will be introduced. 
\subsection{Continuum scaling limit of the transfer matrix\label{3.1}}
In \cite{FPR2} the finite size (order $1/N$) corrections to the lattice 
energies (scaling energies) were computed with standard TBA techniques
that involve a {\em continuum scaling limit} on the lattice model.
This becomes fully apparent when the bulk of the lattice is off the critical 
temperature, $t\neq 0$ ($t=(T-T_c)/T_c$), and the continuum limit 
($N\rightarrow \infty, a\rightarrow 0$, $a$ is the lattice spacing) 
is sensible only if the temperature is also scaled, $t\rightarrow 0$ 
as $N |t|^{\nu}=\mu=const.$ ($\nu=5/4$) \cite{PearceN}.
Physically, this corresponds to a huge magnification of the critical region.
The bulk critical point itself is reached when the dimensionless mass $\mu$
goes to zero: $\mu\rightarrow 0$. 
In the present case, the bulk is at the critical temperature ($\mu=0$)
and we use the boundary perturbation to move away from the 
boundary critical points.

The aim of this section is to show how to define a meaningful 
continuum scaling limit of the lattice transfer matrix. 
Let us recall that in \cite{FPR2}, from the  derivation of the TBA equations,
it was shown that
the following scaling property for the  normalized transfer matrix holds
\begin{equation}\label{scalet}
\hat{\mathbf{t}} (u,\xi)=\lim_{N\rightarrow \infty}
\mathbf{t}^{N}(u+\frac{i}{5}\log N,\xi-\log N)
\end{equation}
and hence a proper 
definition of the continuous transfer matrix (i.e., a transfer matrix
for the continuum theory) is
\begin{equation}\label{conttm}
\hat{\mathbf{D}} (u,\xi)=\lim_{N\rightarrow \infty}\mathbf{D}^{N} 
(u+\frac{i}{5}\log N,\xi-\log N)
\left[ \frac{2\sin \lambda e^{i(u-\frac{\lambda}{2})}}
{N^{1/5}} \right]^{2N}
\end{equation}
where we can explicitly observe the periodicity property 
$\hat{\mathbf{D}} (u,\xi)=\hat{\mathbf{D}}(u+\pi,\xi)$. Since $\mathbf{D}^{N}$ is 
real in the center of each strip (as shown in \cite{FPR2}) and the square of the
factor in brackets is real for $\re(u)=\lambda/2, ~3\lambda$, we deduce that also
$\hat{\mathbf{D}}$ is real in the center of each strip. 

The main motivation to introduce the previous definition is to have an object 
for the continuum theory with the same analytic structure of 
the lattice transfer matrix. Hence, let us analyze such a structure starting
from the string content of the eigenvalues of the double-row transfer matrix. 
A given eigenvalue $D^{N}(u,\xi)$ is characterized by its content of 1-strings. 
It is important to notice that, increasing $N$,  
the number of 1-strings, of 2-strings in the second strip and the relative 
position of 1- and 2-strings does not change. 
Instead, it is observed the appearance of new 2-strings 
close to the real axis in the first strip, moreover eq. (\ref{n1n2}) 
tells us that $n_1$ grows as $N/2$ 
(notice that $N$ grows in steps of 2 so that $n_1$ grows in steps of 1). 
Such new 2-strings push away the remaining zeros from the real axis
and since the extension of the region occupied by them grows as  
$\log N$, then the following limit 
is finite ($v_k^{(j)}$ is a function of $N$): 
\be\label{yscal}
y_k^{(j)}\stackrel{\text{def}}{=}\lim_{N\rightarrow\infty}(5 v_k^{(j)}-\log N).
\ee
In terms of the 
scaled coordinate $y_k^{(j)}$, the real axis located at $\im(u)=0$ is shifted to 
$-i\infty$. From (\ref{conttm}) and (\ref{yscal}) one is immediately led to the 
following observation:
\be
\lim_{N\rightarrow \infty}D^{N} 
(\frac{1}{2}\lambda +\frac{i}{5}(5v_k^{(1)}-\log N)+\frac{i}{5}\log N,
\xi-\log N)=0=
\hat{D}(\frac{1}{2}\lambda +\frac{i}{5}y_k^{(1)},\xi)
\ee
(where a similar equation holds for the zeros in the second strip).
The remarkable consequence of such an equation is that the content of zeros 
of an eigenvalue survives 
the continuum limit, the only difference being the 
infinite number of 2-strings that appear in the first strip,
for $u\rightarrow -i\infty$, because of the growth of $n_1\sim N/2$.
 
In order to show that the limit in (\ref{conttm}) is indeed meaningful, in the 
sense that it is finite, we have to take into account the following ingredients. 
Firstly, the derivation of the TBA equations shows the existence of the limit
in (\ref{scalet}); secondly, examining the normalization 
coefficients $S_2(u,\xi), ~S(u)$ in eq. (\ref{normal}) we have
\bea
\hat{S}_2(u,\xi)&=&\lim_{N\rightarrow\infty} S_2(u+\frac{i}{5}\log N,-\lambda+
\frac{i}{5}(\xi-\log N)) \label{scaleS}\\
& = &\frac{-\sin \!\lambda\, \sin(u+\frac{i}{5}\xi+2\lambda)}
{\sin(u+\frac{i}{5}\xi-2\lambda)\sin(u+\frac{i}{5}\xi+\lambda)}
e^{i(u-\frac{1}{2}\lambda)}  \nonumber \\
\hat{S}(u)&=&\lim_{N\rightarrow\infty} S(u+\frac{i}{5}\log N)
=1
\eea
which are finite quantities. The factor in square brackets in (\ref{normal}) gives:
\be 
\left[ \frac{\sin(u+\frac{i}{5}\log N+2\lambda)\sin \!\lambda}
{\sin(u+\frac{i}{5}\log N+3\lambda)\sin(u+\frac{i}{5}\log N+\lambda)} 
\right]^{2N} \mathop{\sim}_{N \to \infty} 
\left[ \frac{2\sin \lambda e^{i(u-\frac{\lambda}{2})}}
{N^{1/5}} \right]^{2N}
\ee
that exactly cancels the corresponding factor in (\ref{conttm}).
Therefore we get the equation 
\be
\hat{\mathbf{t}} (u,\xi)=\hat{\mathbf{D}} (u,\xi)\hat{S}_2(u,\xi)
\ee
which shows that $\hat{\mathbf{D}} (u,\xi)$ is finite because 
$\hat{\mathbf{t}} (u,\xi)$ and $\hat{S}_2$ are finite. 
It also shows that, in the continuum theory, $\hat{\mathbf{t}}$ and 
$\hat{\mathbf{D}}$ are almost equivalent, the difference being in zeros 
and poles explicitly dependent upon $\xi$ that are introduced in 
$\hat{\mathbf{t}}$ by the factor $\hat{S}_2$. On the other hand, the 
lattice $\mathbf{t}^N$ possesses some non physical 
zeros and poles of order $2N$ contained in the square bracket term of  
(\ref{normal}). Moreover, for a generic $A_p$ ABF model, the zeros have to be
understood as zeros of $D^N$ while their occurrence in $t^N$ 
is much more involved.

Finally, it can be shown that 
the functional equation (\ref{functional}) keeps the same form even 
after the continuum scaling limit
\begin{equation} \label{cfunct}
\hat{\mathbf{t}}(u,\xi) \: \hat{\mathbf{t}}(u+\lambda,\xi)=
1+ \hat{\mathbf{t}}(u+3\lambda,\xi).
\end{equation}
This equation can be solved for the eigenvalues of $\hat{\mathbf{t}}$ 
and the solution is given by the TBA equations obtained in \cite{FPR2}. 

It is important to notice that the normalization of 
$\mathbf{t}^{N}(u,\xi)$ is uniquely fixed
by the requirement that the functional equation holds in the form 
(\ref{functional}), namely no multiplicative or additive constants are 
allowed (in particular, this fixes the relation (\ref{normal}) between 
$\mathbf{D}^{N}(u,\xi)$ and $\mathbf{t}^{N}(u,\xi)$). 
Since such a functional equation remains unchanged after the 
continuum scaling limit, the lattice normalization uniquely fixes
the continuum one. 
This fact will have important consequences for a correct definition of the 
integrals of motion.

\subsection{TBA equations and integrals of motion\label{ss_tba}}
Let us discuss the TBA equations for the eigenvalues of $\hat{\mathbf{t}}$. 
First of all, we define
the functions $\hat{t}_1(x,\xi)$ and $\hat{t}_2(x,\xi)$ which are the eigenvalues 
of $\hat{\mathbf{t}}$ referred to the center of each strip:
\bea
\hat{t}_1(x,\xi) &=&\hat{t}(\frac{\lambda}{2}+\frac{i}{5}x,\xi) 
=  s_1  e^{\displaystyle -\epsilon_1(x)}, \\
\hat{t}_2(x,\xi)&=&\hat{t}(3\lambda+\frac{i}{5}x,\xi)
=  s_2 e^{\displaystyle-\epsilon_2(x)}  .
\eea
where the pseudoenergies $\epsilon_j(x)$ are introduced, and
the quantities $s_j=\pm 1$ play the role of integration constants 
and will be fixed later. Since the functions $\hat{t}_j(x)$ are real for 
real $x$ and satisfy $\hat{t}_j(x)> -1$, we can define the real functions 
$L_j(x)$ as
\be
L_j(x)=\log(1+\hat{t}_j(x,\xi)).  
\ee
In terms of the functions $\hat{t}_1$ and $\hat{t}_2$, the functional equation
(\ref{cfunct}) becomes
\bea
\hat{t}_1 (x+ i \frac{\pi}{2} ) \, \hat{t}_1 (x- i \frac{\pi}{2} )
& = & 1+\hat{t}_2 (x) \nonumber \\
\hat{t}_2 (x+ i \frac{\pi}{2} ) \, \hat{t}_2 (x- i \frac{\pi}{2} )
& = & 1+\hat{t}_1 (x).
\eea
Consequently the TBA equations can be written as 
\newcommand{\integ}[1]{\int_{-\infty}^{+\infty}\frac{dy}{2\pi}\,\frac{#1}
{\cosh(x-y)}}
\bea
\label{tba1}  \hat{t}_1(x,\xi) &=& s_1 \hat{g}_1(x,\xi)
\prod_{k=1}^{m_1} \tanh \frac{y_k^{(1)} -x}{2} \;
\exp \left(\integ{L_2(y)}\right),\\
\label{tba2}  \hat{t}_2(x,\xi) &=& e^{-4 e^{-x}} s_2 \hat{g}_2(x,\xi)
\prod_{k=1}^{m_2} \tanh \frac{y_k^{(2)} -x}{2}\;\exp\left(\integ{L_1(y)}\right)
\eea
and the positions of the zeros are fixed by the quantization conditions
(note the inversion of the indices: $\psi_1$ is for strip 2 and
$\psi_2$ for strip 1):
\bea
\label{quant1}
\psi_2(y_k^{(1)}) &=& n_k^{(1)} \pi
=\left(2(I^{(1)}_k+m_1-k)+1-m_2\right)\pi,   \\[2mm]
\label{quant2}
\psi_1(y_k^{(2)}) &=& n_k^{(2)} \pi
=\left(2(I^{(2)}_k+m_2-k)+1-m_1\right)\pi. 
\eea
The same equations hold for the 2-string locations $z_l^{(j)}$ so, each
time $\psi_j(x)= n \pi$ is satisfied for an integer $n$ with the appropriate
parity\footnote{There is the following constraint: 
$n_k^{(j)}+\frac{s_{3-j}+1}{2}=0 \text{~mod~} 2$.}, $x$ is a 1-string 
or a 2-string in the strip $3\!-\!j$.
\newcommand{\pvint}{\!\rule[-2mm]{0mm}{2mm}_{\scriptscriptstyle PV}
\!\!\int_{-\infty}^{+\infty}}
The explicit expressions for $\psi_j$ are given by:
\begin{eqnarray}
\psi_1(x)  &\equiv& \re \left( -i\,\epsilon_1(x-i\frac{\pi}{2})\right) =
\re \left( i \log(s_1 \hat{t}_1(x-i\frac{\pi}{2},\xi)) \right) \label{psi1}  \\
&=& i\log \hat{g}_1(x-i\frac{\pi}{2},\xi) 
+i\sum_{k=1}^{m_1} \log \tanh \big(\frac{y_k^{(1)} -x}{2}
+i\frac{\pi}{4}\big) -\pvint \frac{dy}{2\pi}\,\frac{L_2(y)}{\sinh(x-y)}, 
\nonumber \\
\psi_2(x)  &\equiv& \re \left( -i\,\epsilon_2(x-i\frac{\pi}{2})\right) =
\re \left( i \log(s_2 \hat{t}_2(x-i\frac{\pi}{2},\xi)) \right) \label{psi2}\\
&=& 4 e^{-x}+ i\log \hat{g}_2(x-i\frac{\pi}{2},\xi) 
+i\sum_{k=1}^{m_2} \log \tanh \big(\frac{y_k^{(2)} -x}{2}
+i\frac{\pi}{4}\big) -\pvint
\frac{dy}{2\pi}\,\frac{L_1(y)}{\sinh(x-y)}\nonumber
\end{eqnarray}
where the integral around the singularity $x=y$ is understood as a
principal value. Moreover, for numerical convenience, 
we take the fundamental
branch for each logarithm so that in general
$\log a + \log b$ cannot be identified with $\log (a b)$.
The energy predicted for each state is given by:
\begin{equation}\label{scalingenergy}
E(\xi)=
\displaystyle \frac{2}{\pi} \sum ^{m_{1}}_{k=1}e^{-y^{(1)}_{k}}-
\int ^{\infty }_{-\infty } \frac{dy}{\pi^2}\,e^{-y} L_2(y)
\end{equation}
and reduces to (\ref{crit}) at both the critical points (compare also with (\ref{leading})).
Since we are interested in the flow $\chi_{1,2}\mapsto \chi_{1,1}$, we have
\be \begin{array}{c}
s_1=s_2=1; \\
\hat{g}_1(x,\xi)=1; \quad
\hat{g}_2(x,\xi)=\displaystyle \tanh\frac{x+\xi}{2}.
\end{array}\ee
It is useful to recall that the functions $\hat{t}_j(x,\xi)$ are eigenvalues of a
commuting family of operators:
\be \label{commtcont}
[\hat{\mathbf{t}}(u,\xi),\hat{\mathbf{t}}(u',\xi)]=0.
\ee
Hence we are in the position to show how the local integrals of motion emerge from
the continuum scaling limit of the transfer matrix $\hat{\mathbf{t}}(u,\xi)$. 
First of all we expand the quantity $\log \hat{t}_1$ 
for $x\rightarrow -\infty$ where $\hat{t}_1$ is given by (\ref{tba1}). 
In order to do so we need the 
following formul\ae  
$$\begin{array}{l}\displaystyle
\log \tanh\frac{y_k^{(1)}-x}{2} =\log \frac{1-e^{-y_k^{(1)}+x}}
{1+e^{-y_k^{(1)}+x}} \\[7mm]\displaystyle
\log \frac{1-t}{1+t}=-2\Big( t+\frac{t^3}{3}+\frac{t^5}{5}+\ldots \Big)
, \qquad |t|<1 \\[5mm] \displaystyle
\frac{1}{\cosh x}=2\sum_{n=1}^{\infty}(-1)^{n-1} e^{(2n-1)x} \, , \qquad x<0.
\end{array}
$$
Hence, taking advantage of the previous results, 
we arrive to the following expressions
for the continuum scaling limit of the logarithm of the transfer matrix 
\begin{eqnarray}
\log \hat{t}_1(x,\xi) &=&-\sum_{n=1}^{\infty} C_n 
I_{2n-1}(\xi) e^{(2n-1)x} , \label{asym}\\
C_n  I_{2n-1}(\xi)&
=&\frac{2}{2n-1}\sum_{k=1}^{m_1}e^{-(2n-1)y_k^{(1)}}
+(-1)^{n}\int_{-\infty}^{+\infty}\frac{dy}{\pi}e^{-(2n-1)y}L_2(y), \label{im}
\end{eqnarray}
where the quantities\footnote{We hope that the symbols $I_k^{(j)}$, 
with $j=1,2$ introduced for the quantum numbers, are not confused with the 
similar symbols $I_{2n-1}$ used for the integrals 
of motion.} $I_{2n-1}$ are the eigenvalues 
of the infinite family of integrals of motion in 
involution $\{\I_{2n-1}\}_n$.
It is important to stress that the TBA calculation provides only the 
value of the product $C_n  I_{2n-1}(\xi)$ as it is clear from eq. 
(\ref{asym}) and (\ref{im}).
The coefficients $C_n$ will be fixed later by imposing 
appropriate normalization conditions (see sect.~\ref{s_comp}).  

As pointed out in sect.~\ref{3.1}, the normalization of $\hat{t}_1$
is fixed once forever on the lattice and survives the continuum scaling limit
hence its largest eigenvalue, being the lattice groundstate, in that limit
is directly mapped to the true vacuum of the continuum field theory.
This forbids the presence of additive constants to $I_{2n-1}$ in (\ref{asym}). 
Such a thing is in agreement with the general 
properties one expects for the above integrals of motion in the continuum 
limit:
since they are given by integrals over local densities, and the local 
densities are components of Lorentz vectors, there is no freedom to 
add any arbitrary scalar contribution to them.
Hence we expect to find, at the fixed points of the massless flow,
the same values for $I_{2n-1}$ given by the conformal field theory
(see next section)
once the proper normalization for the corresponding $C_n$ has been chosen.

As an example, it is interesting to note that 
comparing (\ref{im}) with (\ref{scalingenergy}) we obtain $C_1=\pi$ and 
$I_1(\xi)=E(\xi)$: as expected, the first integral of motion is the energy. 

Now, our purpose is to use the TBA equations to compute the 
eigenvalues of $\I_{2n-1}$ and compare them with their analog from 
the conformal field theory.
As we will see, the two sets of integrals of motion precisely correspond
to one another.

\section{Integrals of motion in conformal field theory\label{s_IM}}
Conformal field theories are integrable, in the sense that they possesses
an infinite number of mutually commuting integrals of motion.
Their construction was given in \cite{SY} by canonically quantizing 
the family of classical integrals of motion of the 
Modified Korteweg--de Vries equation (that also coincide with the 
integrals of motion of the classical sine-Gordon equation).

A general expression for such integrals of motion is not known but they 
can all be obtained in a constructive way. They are local expressions 
(polynomials) in the Virasoro generators and   
the first few of them are given by\footnote{It is convenient 
to indicate each integral of motion with $\I_{2n-1}$ and the 
corresponding eigenvalues with $I_{2n-1}$.} (see \cite{blz94} for their explicit expressions)
\begin{eqnarray}
\I_1&=&L_0 -\frac{c}{24} \nonumber \\
\I_3&=&2\sum_{n=1}^{\infty}L_{-n}L_n +L_0^2-\frac{2+c}{12}L_0
+ \frac{c(5c+22)}{2880}
\label{elencoim}\\
\I_5&=&\sum_{m,n,p}:\!L_n L_m L_p\!:\delta_{0,m+n+p}+
\frac{3}{2}\sum_{n=1}^{\infty}L_{1-2n}L_{2n-1} \nonumber \\
& &+\sum_{n=1}^{\infty}\left(\frac{11+c}{6}n^2-\frac{c}{4}-1 \right)L_{-n}L_n 
-\frac{4+c}{8} L_0^2 \nonumber \\ 
& & +\frac{(2+c)(20+3c)}{576}L_0 - \frac{c(3c+14)(7c+68)}{290304}  \nonumber
\end{eqnarray}
(the symbol $:~:$ indicates normal order, obtained by arranging the 
operators $L_n$ in an increasing sequence with respect to their indices).

Their simultaneous diagonalization is, in principle, a straightforward
but lengthy calculation of linear algebra. 
Actually, the rapidly growing complexity
of the involved expressions makes the computation possible 
for the first few levels only. This becomes clear if we look at 
the characters of the Virasoro algebra:
$$\chi_{r,s}(q)=q^{-\frac{c}{24}}~\text{Tr}~q^{L_0}
$$ 
where the trace is taken in the irreducible module of the highest weight
state $(r,s)$.
For example, for the TIM one has:
\bea
\chi_{1,1}(q)&=&q^{-\frac{c}{24}}(1+q^2+q^3+2q^4+2q^5+4q^6+4q^7+7q^8+\ldots )\\
\chi_{1,2}(q)&=&q^{-\frac{c}{24}+\Delta_{1,2}} 
(1+q+q^2+2q^3+3q^4+4q^5+6q^6+8q^7+\ldots).
\eea
In the previous expression, the coefficient $\alpha_n$ in the monomial 
$\alpha_n q^n$ gives the degeneracy of the subspace at level $n$.
At a given level of truncation, the expression for the vacuum sector 
$\chi_{1,1}(q)$ holds for all the 
unitary minimal models except the $c=1/2$ Ising model. 
Up to level 3 there is no degeneracy and the eigenstate is given by the 
unique linearly independent vector. At level 4 and 5 the double degeneracy 
leads to an algebraic quadratic equation that can be easily solved. 
At level 6 the degeneracy of order 4 leads to a quartic algebraic equation
and, in general, one realizes that the complexity in managing this problem 
rapidly grows with the degeneracy of the level. 
More details on this topic can be found in Tables~A1 and A2 of the 
Appendix where we present a list of eigenstates/eigenvalues for the 
vacuum sector of all unitary minimal models (Table~A1) and  for the sector 
$(1,2)$ of TIM (Table~A2).
From Tables~A1 and A2, we notice that $\I_3$ alone is enough to
completely remove the degeneracy, at least up to level 6. 

The list of integrals of motion discussed here survives the off 
critical perturbation by $\varphi_{1,3}$ \cite{zam}. 
More precisely, off critical involutive integrals of motion $I_{2n-1}(\lambda)$
exist for the perturbed CFT\footnote{With $\mathcal{M}_{p,q}$ we refer to the
minimal CFT with central charge $c=1-\frac{6(p-q)^2}{pq}$.} 
$ \mathcal{M}_{p,q}+\lambda \varphi_{1,3}$
and at criticality they reduce to the expression (\ref{elencoim}).


\section{TBA data versus CFT\label{s_comp}}
As anticipated in the introduction, we are interested in studying 
the model both at criticality and along the boundary flow 
\mbox{$\chi_{1,2}\mapsto\chi_{1,1}$} generated by the 
$\varphi_{1,3}$ integrable perturbation of TIM. Let us first describe the 
sector (1,1) of the theory, i.e. the IR critical point. 
Let us recall that TBA computations provide the product 
$C_n  I_{2n-1}(\xi)$, see eq.~(\ref{im}). The constants $C_n$ are 
precisely introduced to make contact with the integrals of motion of the CFT.
As already pointed out at the end of sect.~\ref{ss_tba},
the first of them is exactly know, $C_1=\pi$.
Since we are interested in $\I_3$ and $\I_5$ we can numerically 
fix the corresponding
constants $C_2$ and $C_3$ using the vacuum state\footnote{Actually, 
one reaches higher numerical precision using the first excited state and
so we did.} 
\bea 
 C_2 =2.1838434, \ \ \ \ \ 
 C_3 =3.7555032.
\eea
The general expression was computed in \cite{blz94} and reads\footnote{We 
would like to thanks the referee of Nucl. Phys. B  for having pointed out
to us this equation.}:
\be
C_n=2^{2-n}\ 3^{1 - 2 n}\ 5^{1 - n}\ \frac{(10 n-7)!!}{n!\, (4 n-2)!} \ \pi .
\ee
In particular, this gives: 
$C_1=\pi$, $C_2=\frac{1001}{1440}\, \pi\sim 2.1838432 $
and $C_3=\frac{7436429}{6220800}\, \pi \sim 3.7555026$.
\openin1=dati_I3_11cft.tex
\openin2=dati_I3_11tba.tex
\openin3=dati_I5_11cft.tex
\openin4=dati_I5_11tba.tex
\newcommand\ra{\read1 to \datoa \datoa}
\newcommand\rb{\read2 to \datob \datob}
\newcommand\rc{\read3 to \datoc \datoc}
\newcommand\rd{\read4 to \datod \datod}
\begin{table}
\caption{\label{tvac}Comparison between the eigenvalues of $\I_3,\ \I_5$ from 
CFT and from TBA in the vacuum sector of TIM.
The left column indicates the 
level degeneracy (l.d.) predicted by the character of the Virasoro algebra.}
\vspace{3mm}$$\hspace*{-11.5mm}
\begin{array}{|@{~~}l@{\hspace{-1mm}}r@{~\longleftrightarrow~}l@{~~~~}l@{~~~~}l@{~~~~}l@{~~~~}l|}
\hline \rule{0mm}{6mm}
\text{l.d.} & \multicolumn{2}{l}{\hspace{10mm}\text{lattice-conformal dictionary}}
 &\ra&\rb&\rc&\rd\\[2mm] \hline 
\rule{0mm}{8mm} 1 & (~) &\v  &\ra&\rb&\!\!\!\!\rc&\!\!\!\!\rd \\[10mm]
1q^2 & (00) & L_{-2}\v &\ra&\rb&\rc&\rd \\[10mm]
1q^3 & (10) & L_{-3}\v &\ra&\rb&\rc&\rd \\[10mm]
2q^4 & (20) & 3(\frac{4+\sqrt{151}}{5} L_{-4}+2\, L_{-2}^2)\v &\ra&\rb&\rc&\rd \\[4mm]
     & (11) & 3(\frac{4-\sqrt{151}}{5} L_{-4}+2\, L_{-2}^2)\v &\ra&\rb&\rc&\rd  \\[10mm]
2q^5 & (30) &(\frac{7+\sqrt{1345}}{2} L_{-5}+20\, L_{-3}L_{-2})\v &\ra&\rb&\rc&\rd \\[4mm]
     & (21) &(\frac{7-\sqrt{1345}}{2} L_{-5}+20\, L_{-3}L_{-2})\v &\ra&\rb&\rc&\rd \\[10mm]
4q^6 & (40) & (11.124748\, L_{-6}+9.6451291\, L_{-4}L_{-2}  &\ra&\rb&\rc&\rd \\ 
 & \multicolumn{2}{l}{\hspace{27mm}+4.4320186\, L_{-3}^2+ L_{-2}^3)\v } &&&& \\[4mm]
 & (31) &(-4.9655743\, L_{-6} + 2.3354391\, L_{-4}L_{-2}   &\ra&\rb&\rc&\rd \\
 & \multicolumn{2}{l}{\hspace{27mm}+0.71473858\, L_{-3}^2+ L_{-2})\v} &&&& \\[4mm]
 & (22) &(0.66457527\, L_{-6} - 1.2909210\, L_{-4}L_{-2} &\ra&\rb&\rc&\rd\\
 & \multicolumn{2}{l}{\hspace{27mm}-1.2605013\, L_{-3}^2+L_{-2}^3)\v} &&&& \\[4mm]
 & (0000|00) &(-1.6612491\, L_{-6} - 4.0646472\, L_{-4}L_{-2} &\ra&\rb&\rc&\rd \\
 & \multicolumn{2}{l}{\hspace{27mm}+ 1.4118691\, L_{-3}^2 + L_{-2}^3)\v} &&&&\\[3mm] \hline
\end{array}$$
\end{table}
\closein1\closein2\closein3\closein4
\openin5=dati_I3_12cft.tex
\openin6=dati_I3_12tba.tex
\openin7=dati_I5_12cft.tex
\openin8=dati_I5_12tba.tex
\newcommand\ree{\read5 to \datoe \datoe}
\newcommand\rf{\read6 to \datof \datof}
\newcommand\rg{\read7 to \datog \datog}
\newcommand\rh{\read8 to \datoh \datoh}
\begin{table}
\caption{\label{t12}Comparison between the eigenvalues of $\I_3,\ \I_5$ from 
CFT and from TBA in the $(1,2)$ sector of TIM.
The left column indicates the 
level degeneracy (l.d.) predicted by the character of the Virasoro algebra.}
\vspace{3mm}$$\hspace*{-11mm}
\begin{array}{|@{~~}l@{\hspace{-1mm}}r@{~\longleftrightarrow~}l@{~~~~~~~~}r@{~~~~}r@{~~~~}r@{~~~~}r@{~~}|}
\hline \rule{0mm}{6mm}
\text{l.d.} & \multicolumn{2}{l}{\hspace{10mm}\text{lattice-conformal dictionary}} &\ree&\rf&\rg&\rh \\[2mm] \hline 
\rule{0mm}{8mm} 1 & (0)_{+}  & \gs & \ree &\!\!\!\! \rf & \rg & \rh \\[10mm]
1q^1 & (0)_{-} &L_{-1}\gs& \ree&\rf&\rg&\rh \\[10mm]
1q^2 & (1)_{-} &L_{-2}\gs&  \ree&\rf&\rg&\rh \\[10mm]
2q^3 & (2)_{-} &\bigl( 16\, L_{-3} + (3+\sqrt{649}) L_{-2}L_{-1} \bigr)\gs &  \ree&\rf&\rg&\rh\\[4mm]
     & (000)_{+} &\bigl( 16\, L_{-3} + (3-\sqrt{649}) L_{-2}L_{-1} \bigr)\gs &  \ree&\rf&\rg&\rh\\[10mm]
3q^4 & (3)_{-} & (L_{-4} + 2.1916731\, L_{-3}L_{-1}
     &  \ree&\rf&\rg&\rh \\
     & \multicolumn{2}{l}{\hspace{26mm}+1.2645819\, L_{-2}^2)\gs }& & & & \\[4mm]
     & (100)_{+} & (L_{-4}- 0.50977440\, L_{-3} L_{-1} 
     &  \ree&\rf&\rg&\rh \\
     & \multicolumn{2}{l}{\hspace{26mm}-0.42935709\, L_{-2}^2)\gs  }& & & & \\[4mm]
     & (000|00)_{-} &   (L_{-4}+2.5725923\,  L_{-3}L_{-1}
     &  \ree&\rf&\rg&\rh \\
     & \multicolumn{2}{l}{\hspace{26mm}-1.7645661\, L_{-2}^2)\gs }& & & & \\[10mm]
4q^5 & (4)_{-} & (L_{-5}+1.6541786\, L_{-4}L_{-1}
     &  \ree&\rf&\rg&\rh  \\
     & \multicolumn{4}{l}{\hspace{26mm}+1.3195562\, L_{-3} L_{-2}+0.52470182\, L_{-2}^2 L_{-1})\gs}  & & \\[4mm]
     & (200)_{+} & (L_{-5} - 0.29373082\, L_{-4}L_{-1} 
     &  \ree&\rf&\rg&\rh \\
     & \multicolumn{4}{l}{\hspace{26mm}-0.012595542\, L_{-3}L_{-2}-0.45966766\, L_{-2}^2 L_{-1})\gs}  & & \\[4mm]
     & (110)_{+} & (L_{-5}+0.19041567\, L_{-4} L_{-1}
     &  \ree&\rf&\rg&\rh \\
     & \multicolumn{4}{l}{\hspace{26mm}-2.9757621\, L_{-3}L_{-2}+1.8161567\, L_{-2}^2 L_{-1})\gs}  & & \\[4mm]
     & (100|00)_{-} & (L_{-5}+4.9698982\, L_{-4} L_{-1}
     &  \ree&\rf&\rg&\rh \\
     & \multicolumn{4}{l}{\hspace{26mm}-1.1753145\, L_{-3}L_{-2}-1.4399411\, L_{-2}^2 L_{-1})\gs}  & & \\[3mm]
\hline
\end{array}
$$ \end{table}
After that, we are ready to turn the crank of TBA equations in order to
obtain the other eigenvalues of the operators $\I_{3}$ and $\I_{5}$.
The comparison between them and the corresponding eigenvalues coming 
from CFT can be found in Table~\ref{tvac}. 
We can immediately notice that the agreement
is quite remarkable: the solutions of the TBA equations are given with seven
significant digits of precision and perfectly match the CFT results. \\
Hence we can draw the following conclusions: firstly, the expansion
postulated in (\ref{asym}) is the suitable one for the computation of
the eigenvalues of the integrals of motion in the lattice model (in the
continuum limit) and shed some light upon the connection between the
integrable structure of the CFT and the lattice model;
secondly, such a matching between the eigenvalues 
allows to conjecture a one-to-one correspondence between the states of the
CFT and the corresponding states defined within the lattice model.     \\  
Few remarks are in order with respect to the latter issue. Even if we are able
to build the correspondence of states level by level, we are not able to
write down explicit, closed expressions for the eigenvalues $I_{2n-1}$ at criticality as it
was done for the energy (see (\ref{crit}) and more complete expressions in \cite{FPR2}).
Notwithstanding this, numerical evidence strongly supports the conjecture that the
state-by-state correspondence is exact giving a 
true \emph{lattice-conformal dictionary}.

Let us study the (1,2) sector of TIM, as in Table~\ref{t12}. 
As expected we get a similar result as
before, confirming the above conjectured correspondence 
(it is useful to stress that the constants $C_n$ are fixed
once forever, because they are computed on the true vacuum that is independent 
of $\xi$).

\begin{figure}\begin{center}\setlength{\unitlength}{1mm}
\begin{picture}(134,180)
\put(0,0){\includegraphics[width=0.8\linewidth]{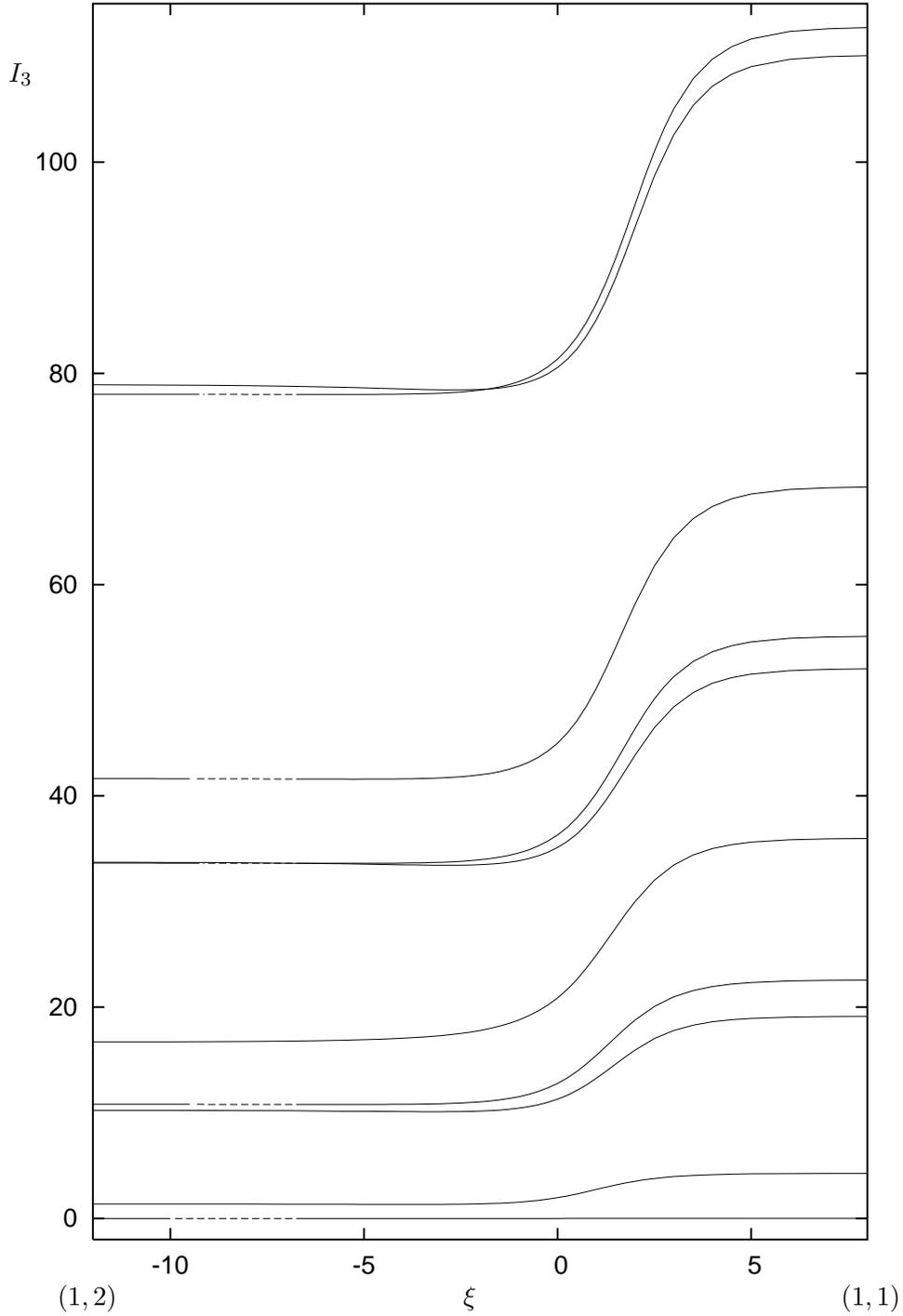}}
\put(10,-2){$(1,2)$}\put(68,-2){$\xi$}\put(122,-2){$(1,1)$}
\put(3,173){$I_3$}
\end{picture}\end{center}
\caption{Behaviour of the eigenvalues of the integral of motion $\I_3(\xi)$ 
along the renormalization group flow. We point out that, in the UV region, 
there is no degeneracy of states; what seems a degeneracy is 
a graphical effect because of the vicinity of the eigenvalues (compare with 
Section~\ref{s_comp}). 
\label{plotI3}}
\end{figure}

Finally, we can explore by means of TBA equations what happens to the 
integrals of motion when the boundary $\varphi_{1,3}$ perturbation is switched on
(see \cite{Fioravanti:2003kx} for the case of a massive perturbation in the context of \cite{Fioravanti:2002sq}).
Since off-critical expressions for the integrals of motion are not available we cannot
make any comparison, but we can just follow the evolution along the whole boundary
flow $\chi_{1,2}\mapsto\chi_{1,1}$. \\
In Fig.~\ref{plotI3} the first few eigenvalues of $\I_{3}$ are plotted (against $\xi$): 
we can notice the absence of degeneracy and the clear correspondence between the
states belonging to the UV and IR fixed points. Such a correspondence has been explicitly shown in Table~\ref{t_s1211}.

The method to solve numerically the TBA equations was 
widely discussed in \cite{FPR2}. 
The computation of the integrals of motion 
does not require any new technique: it simply uses the previously obtained 
solution in the expressions (\ref{im}).  

\begin{table}\caption{\label{t_s1211} Flow $\chi_{1,1} \mapsto \chi_{1,2}$ 
(reverse of the physical flow).
We present the explicit mapping of states from IR to UV. 
Here $n^{\text{IR}},~ n^{\text{UV}}$ are the excitation levels
above the ground states, respectively $\Delta=0$ and $\Delta=1/10$.}
\begin{center}
\begin{tabular}{|c|r@{$\:\:\mapsto\:\:$}l|c||c|r
@{$\:\:\mapsto\:\:$}l|c|}
\hline
$n^{\text{IR}}$\rule[-4mm]{0mm}{10mm}& \multicolumn{2}{c|}{Mapping of states} & $n^{\text{UV}}$ &
   $n^{\text{IR}}$ & \multicolumn{2}{c|}{Mapping of states} & $n^{\text{IR}}$\\
\hline
0\rule[-1mm]{0mm}{6mm} & $(\,)$ & $(0)_{+}$ & 0 &
             5 & $(2\,1)$ & $(1\,0\,0)_{+}$ & 4 \\        
2 & \hspace{2mm}$(0\,0)$ & $ (0)_{-}$ & 1 &
             6 & $(4\,0)$ & $(4)_{-}$ & 5\\ 
3 & $(1\,0)$ & $(1)_{-}$ & 2 &
             6 & $(3\,1)$ & $(2\,0\,0)_{+}$ & 5 \\
4 & $(2\,0)$ & $(2)_{-}$ & 3 &
             6 & $(2\,2)$ & $(1\,1\,0)_{+}$ & 5 \\
4 & $(1\,1)$ & $(0\,0\,0)_{+}$ & 3 &
             6 & $(0\,0\,0\,0|0\,0)$ & $(0\,0\,0|0\,0)_{-}$ & 4 \\
5 & $(3\,0)$ & $(3)_{-}$ & 4 & & \multicolumn{2}{c|}{~}  &\\[1mm]
\hline
\end{tabular}\end{center}\end{table}
\newpage

\section{Continuous transfer matrix from CFT\label{s_cont}}
In the present section we will discuss the relation between the so called 
{\bf T}-operators introduced by Bazhanov, Lukyanov and Zamolodchikov (BLZ) in 
\cite{blz94}, and the continuum limit of the lattice transfer matrix 
$\mathbf{t}^{(N)}$ (\ref{scalet}).

In the series of papers \cite{blz94}, the authors took advantage of 
the integrable 
structure of Conformal Field Theories building up the quantum version of the monodromy matrices of KdV theory. 

A crucial role in their construction is played by the "{\bf T}-operators", which are the trace of such quantum monodromy matrices.
As shown by BLZ, as a consequence of the Quantum Yang-Baxter equations the {\bf T}-operators form an infinite set of operators in involution
\be
[ \textrm{\bf T}_j(\nu),\textrm{\bf T}_{j'}(\mu)]=0
\ee
where $\nu$, $\mu$ are spectral parameters and the indexes $j$, $j'$ label the corresponding representations of $U_q(sl(2))$.
Such a quantum group structure has its root in the Feigin-Fuchs construction related to the quantum version of the monodromy matrices. Here it is important to recall that since the stress energy tensor is defined as 
\be
-\beta^2 T(u)= :\phi'(u)^2:+(1-\beta^2)\phi''(u)+\frac{\beta^2}{24}
\ee
where $\phi$ is the compactified bosonic field (see \cite{blz94} for details),
then the relation between the central charge $c$ of the CFT and the parameter $\beta$ is
\be
c = 13-6(\beta^2+\beta^{-2})
\ee
and the deformation parameter $q$ of the quantum algebra is related to $\beta$ as
\be
q \ = \ e^{i \pi \beta^2 }.
\ee 
BLZ analyzed in detail the properties of the operators $ \textrm{\bf T}_j(\nu)$ establishing, in particular, that they satisfy the following (infinite) set of functional relations
\be
\label{blzfunct}
\textrm{\bf T}_j(q^{\frac12} \nu)  \textrm{\bf T}_j(q^{-\frac12} \nu)=1+  \textrm{\bf T}_{j+\frac12}(\nu) \textrm{\bf T}_{j-\frac12}(\nu)
\ee   
showing that at particular values of the parameter $\beta$ a truncation of such a system of functional relations occurs. In \cite{blz94} this phenomenon was shown explicitly for the non-unitary series of minimal models $\mathcal{M}_{2,2n+3}$, giving rise to a set of equations which is equivalent to the $Y$-system of the TBA equations for the massless $S$-matrix associated to the previous CFT's. As pointed out in \cite{blz94}, the functional relations (\ref{blzfunct}) also hold in the case of the $\varphi_{1,3}$ perturbation of $\mathcal{M}_{2,2n+3}$. Taking advantage of this fact in \cite{blzxx} the solutions of the TBA equations for the excited states of the Yang-Lee model ($\mathcal{M}_{2,5}$) with $\varphi_{1,3}$ perturbation were found. \\    
As anticipated in \cite{blz94}, such an analysis can be repeated \textit{mutatis mutandis} for the series of minimal unitary models $\mathcal{M}_{p,p+1}$ where $p=3,4,\dots$ is an integer which labels their central charges
\be
c = 1 - \frac{6}{p(p+1)}.
\ee 
It can be shown that in this case the truncation restricts the range of variation of the index $j$ of the $ \textrm{\bf T}_j(\nu)$ operators to the subset $j\in \{0,\frac12,1,\dots,\frac{p-1}{2} \}$ leading also to the following symmetry relation
\be
\textrm{\bf T}_{\frac{p-1}{2}-j}(\nu) =  \textrm{\bf T}_j(\nu).
\ee 
If we refer to $T_j(\nu)$ as the eigenvalues\footnote{Our notation for the eigenvalues of $\textrm{\bf T}$ differs from that of \cite{blz94} in order to
avoid confusion with those of the normalized transfer matrix $\hat{\mathbf{t}}$.}
 of $ \textrm{\bf T}_j(\nu)$, then the following relations can be written
\bea
\label{eqsys}
T_j(q^{\frac12} \nu) T_j(q^{-\frac12} \nu) & =  &1+ T_{j+\frac12}(\nu) T_{j-\frac12}(\nu) \nonumber \\
T_{\frac{p-1}{2}-j}(\nu)   =   T_j(\nu);  & &    T_{\frac{p-1}{2}}(\nu)  =  T_0(\nu)=1. 
\eea
Inspired by the results of $\cite{blz94,rava,kp}$, we perform the substitution 
$(a=2j)$
\be
\Upsilon_{a}(\theta)=T_{\frac{a}{2}+\frac12}(\nu) T_{\frac{a}{2}-\frac12}(\nu); \ \ \ \ \nu=e^{\beta^2\theta} 
\ee
which allows to cast eq. (\ref{eqsys}) in the form
\bea
\label{blzY}
\Upsilon_a (\theta + \frac{i \pi}{2})\Upsilon_a (\theta - \frac{i \pi}{2}) & = & (1+\Upsilon_{a+1}(\theta))(1+\Upsilon_{a-1}(\theta)) \nonumber \\
\Upsilon_{p-1-a}(\theta)   =   \Upsilon_a(\theta);  & &    \Upsilon_{p-1}(\theta)  =  \Upsilon_0(\theta)=0. 
\eea
where $a \in \{ 0,1,2,\dots,p-1\}$.
Now we are in the position to make contact with the system of TBA functional equations for the $\varphi_{1,3}$ perturbation of the theories $\mathcal{M}_{p,p+1}$, obtained from a standard scattering approach \cite{zamoflux}. Such equations can be written as 
\bea
\label{tbaY}
Y_a (\theta + \frac{i \pi}{2})Y_a (\theta - \frac{i \pi}{2})  =  \prod_{b=0}^{p-1} (1+Y_b(\theta))^{\ell_{ab}} =(1+Y_{a+1}(\theta))(1+Y_{a-1}(\theta))
\eea  
where $\ell_{ab}=\delta_{a+1,b}+\delta_{a-1,b}$ is the incidence matrix of the Dynkin diagram of $A_{p-2}$ and  $Y_0(\theta)=Y_{p-1}(\theta)=0$.
The equivalence between (\ref{blzY}) and (\ref{tbaY}) can be immediately recognized.   

The same equations were first obtained by Kl\"umper and Pearce \cite{kp} for the integrable RSOS fusion hierarchies. 
This reveals that there are identical structures coming from the 
integrable lattice and from the minimal models of conformal field theory.
In both the cases a transfer matrix can be defined such that it
obeys the same functional equations. The connection between the BLZ formalism 
and the transfer matrix formalism of \cite{kp} is given by the 
conjectured correspondence: 
\be\label{corrisp}
\Upsilon_a(\theta)\longleftrightarrow\hat{t}_{a}\left(-\frac{i}{p+1}\theta\right)
\ee
which means that they are equal modulo some lattice-continuum 
normalization constants. 
Such a conjecture has been numerically tested (at criticality)
 in sect.~\ref{s_comp} 
for the tricritical Ising model. In particular, we checked that 
the two terms in (\ref{corrisp}) contain the same integrals of motion 
eigenvalues, under the expansion defined in (\ref{asym}). 

The previous reasoning suggests that a suitable prescription to find the solutions to eqs.~(\ref{blzY})
can be that of determine the analytic structure of the $\Upsilon_a$ 
by means of the
pattern of zeros given by the continuum scaling limit of the lattice model.

\section{Conclusions}
The most important result we obtained in this paper is the state-by-state 
correspondence between the $A_4$ ABF lattice model and the integrable QFT 
defined as the boundary $\varphi_{1,3}$ perturbation of the $c=7/10$ minimal unitary
CFT describing the universality class of TIM. \\
Such a result can be added to the previous {\em corpus} of known information and results 
from both the $A_4$ lattice model and the corresponding 
$c=7/10$ CFT possibly perturbed by the boundary operator $\varphi_{1,3}$. As an example
we can mention that in \cite{OPW} it was found that the
limit of the lattice model exactly reproduces the Virasoro characters of the 
corresponding $c=7/10$ minimal unitary CFT.
It is tempting to suggest that the presence of common structures (functional equations,
characters) implies a strong (algebraic) identification between the $A_4$ lattice model at
the scaling limit and the conformal theory. 

Let us remember that the crucial ingredient which allowed us to establish
the correspondence of states is the infinite set of integrals of motion coming
from integrability. By means of them, an unambiguous procedure to fix the 
state-by-state identification has been given, and supported by a clear numerical
evidence, at both the UV and IR critical 
points (the sectors $(1,2)$ and $(1,1)$ respectively) establishing a true
lattice-conformal dictionary (see Section ~\ref{s_comp}). 

Furthermore, we followed the behaviour of the IM along the boundary flow. Since 
it is known that the integrable structure of $\{\I_{2n-1}(\xi)\}_n$ survives off 
criticality, we can see, using the description of the flow given in \cite{FPR2} and the 
results of the present paper, that there is a smooth flow of the states which allows 
to see which state in the sector $(1,2)$ flows to which state in the sector $(1,1)$.     

It is important to stress that our constructive method to build the lattice-conformal dictionary 
essentially works level-by-level with an unambiguous procedure, however at present we are not 
able to see whether it is possible to write general expressions or not. Such a situation seems
to resemble that of the computation of the expressions for the operators $\I_{2n-1}$: 
they can be computed one after the other but general formulas are not available. 

Another important result regards the connection (see sect.~\ref{s_cont}) 
between the ``{\bf T}-operators'' defined
in \cite{blz94} with the transfer matrix $\hat{\mathbf{t}}$ given in 
sect.~\ref{3.1}.

Let us conclude with some remarks about possible developments and extensions of the
present work. Firstly, the existence of a lattice-conformal dictionary seems to
suggest the possibility to define a ``lattice Virasoro algebra'' at least within the
lattice ABF models (interesting approaches to a lattice Virasoro algebra can
be found in \cite{Koo:1993wz,Itoyama:1986ad}). 
At present such a construction is prevented by the fact that 
we are not able to write down general expressions for the state-by-state correspondence.
Secondly, we expect one can get a clear idea of what quasi-particle states
are at and off criticality. 

Finally, one can think to generalize the procedure to all the cases where a TBA
is known for excited states.

\section*{Acknowledgments}
We would like to thank P. Pearce for useful discussions and for the
careful reading of the manuscript. 
This work was partially supported by the
European Commission TMR programme HPRN-CT-2002-00325 (EUCLID) and by 
the COFIN ``Teoria dei Campi, Meccanica 
Statistica e Sistemi Elettronici''.

\newsavebox{\tabella}
\savebox{\tabella}{\begin{minipage}{23cm}{\vspace*{-2mm}
{\Large \textbf{Appendix}}\\[12mm]
Table~A1. List of common eigenstates and eigenvalues of $\I_3,\ \I_5$
in the vacuum sector of  minimal models. The eigenstates are 
orthogonal. The first column is the level degeneracy (l.d.) as given by each 
single monomial appearing in the character expression.\\[6mm]
$$\hspace*{-1mm}\begin{array}{r@{~~~}l@{~~~~}l@{~~~~}l}
\text{l.d.} & \text{eigenstate} & I_3 & I_5 \\[7mm] 
1 & \v & \frac{c(22+5c)}{2880} 
       & \frac{-c(14+3c)(68+7c)}{290304} \\[7mm]
1q^2 & L_{-2} \v & \frac{(480+c)(22+5c)}{2880}
     & \frac{(68+7c)(26208+9058c-3c^2)}{290304} \\[7mm]
1q^3 & L_{-3}\v& \frac{\left(2160+c\right) \left(22+5c \right)}{2880} 
     & \frac{(68+7c)(341712+104314c-3c^2)}{290304} \\[7mm]
2q^4 & \Bigl( (1+2c+\sqrt{37+22c+4c^2}) L_{-4} +6 L_{-2}^2 \Bigr)\v 
     & \frac{96000+16342 c+5 c^2+5760 \sqrt{37+22 c+4 c^2}}{2880}
 &\frac{(68+7c)(1141056+260050c-3c^2+120960\sqrt{37+22c+4c^2})}{290304}\\[2mm]
     & \Bigl((1+2c-\sqrt{37+22c+4c^2}) L_{-4} +6 L_{-2}^2 \Bigr)\v 
       & \frac{96000+16342 c+5 c^2-5760\sqrt{37+22 c+4 c^2}}{2880}
 &\frac{(68+7c)(1141056+260050c-3c^2-120960\sqrt{37+22c+4c^2})}{290304}\\[7mm] 
2q^5& \Bigl((5c+\sqrt{5(48+24c+5c^2)}) L_{-5} +20 L_{-3}L_{-2}\Bigr) \v  
       & \frac{213600+34822c+5c^2+4320\sqrt{5(48+24c+5c^2)}}{2880}
       &\frac{(68+7c)\left(4359600+869386c-3c^2+151200
        \sqrt{5(48+24c+5c^2)}\right)}{290304}\\[2mm]
    & \Bigl((5c-\sqrt{5(48+24c+5c^2)}) L_{-5} +20 L_{-3}L_{-2}\Bigr) \v  
       & \frac{213600+34822c+5c^2-4320\sqrt{5(48+24c+5c^2)}}{2880}
       &\frac{(68+7c)\left(4359600+869386c-3c^2-151200
        \sqrt{5(48+24c+5c^2)}\right)}{290304}\\[7mm]
4q^6& \bigl(\alpha(\beta_i)L_{-6}+\beta_i L_{-4}L_{-2}+\gamma(\beta_i)
   L_{-3}^2+L_{-2}^3\bigr)\v & 12\beta_i+83+\frac{3611}{1440}c+\frac{1}{576}c^2
      & \frac{1}{50-5c+12\beta_i} \Big[\frac{310805}{6}\!-\!\frac{7746725c}{2592}
      \!-\!\frac{20580905c^2}{72576}\!-\!\frac{238025c^3}{72576}\!+\!
      \frac{5c^4}{13824} \\[2mm]
      & i=1,\ldots,4 &  & +\!\big(\frac{87403}{3}\!-\!\frac{212741c}{432}\!
      -\!\frac{560095c^2}{12096}\!-\!\frac{c^3}{1152}\big) \!\beta_i\!
      +\!80(56+c){\beta_i}^2 \!+\!120 {\beta_i}^3 \Big]\\[13mm]
\multicolumn{4}{l}{\hspace*{-1.5mm}\text{We used: }\alpha(\beta)=
\frac{72 \beta^3-6(13c+8)\beta^2+
2(10c^2-41c-1040)\beta+3(15c^2+122c-776)}{18(12\beta-5(c-10))}, \quad
\gamma(\beta)=\frac{6\beta^2-2(2c-11)\beta-9(c+2)}{12\beta-5(c-10)};
\quad\beta_i\text{ are the 4 solutions of the following quartic equation:}} 
\\[3mm]
\multicolumn{4}{l}{\hspace*{-1.5mm} 216\beta^4-54(15c+16)
\beta^3+18(38c^2-39c-428)\beta^2
-(160c^3-1509c^2-10998c-4096)\beta-12(30c^3+70c^2-923c-1526)=0}
\end{array} 
$$}\end{minipage}} 
\openin9=dati_I3_12cftexact.tex
\openin10=dati_I5_12cftexact.tex
\newcommand\ReadLine{\read9 to \datoi \datoi}
\newcommand\ReadLineb{\read10 to \datoj \datoj}
\newsavebox{\tabe}
\savebox{\tabe}{\begin{minipage}{23.5cm}{\vspace*{-4mm}Table~A2.
List of common eigenstates and eigenvalues of $\I_3,\ \I_5$
in the (1,2) sector of TIM, $\Delta=1/10$. The eigenstates are 
orthogonal. The first column is the level degeneracy (l.d.) as given by each 
single monomial appearing in the character expression.}\\[8mm]
$$\hspace*{-1mm}\begin{array}{r@{~~~}l@{~~~~}l@{~~~~}l}
\text{l.d.} & \text{eigenstate} & \ReadLine &  \ReadLineb \\[7mm] 
1 & \gs & \ReadLine &  \ReadLineb  \\[7mm]
1q^1 & L_{-1} \gs & \ReadLine &  \ReadLineb \\[7mm]
1q^2 & L_{-2} \gs & \ReadLine &  \ReadLineb \\[7mm]
2q^3 & \Bigl( 16 L_{-3} + (3+\sqrt{649}) L_{-2}L_{-1} \Bigr)\gs & \ReadLine &  \ReadLineb \\[2mm]
     & \Bigl( 16 L_{-3} + (3-\sqrt{649}) L_{-2}L_{-1} \Bigr)\gs  & \ReadLine &  \ReadLineb\\[7mm]
3q^4& \Bigl( L_{-4} +\alpha_i L_{-3}L_{-1} +\beta(\alpha_i) L_{-2}^2 \Bigr)\gs 
     & \ReadLine &  \ReadLineb  \\[2mm]
     & i=1,\ldots,3 \\[7mm]
4q^5& \bigl( L_{-5} +\gamma_i L_{-4}L_{-1} +\eta(\gamma_i,\kappa_i)
 L_{-3}L_{-2}+\kappa_i L_{-2}^2 L_{-1}\bigr) \gs  & \ReadLine &  \ReadLineb \\[2mm]
     & i=1,\ldots,4 \\[13mm]
\multicolumn{4}{l}{\hspace*{-1.5mm}\text{We used: } \beta(\alpha)=
  \frac{-120 +21\alpha+18\alpha^2}{15(16-7a)}; \quad\alpha_i 
  \text{ are the 3 solutions of the following cubic equation: } 
   334\alpha^3-1421\alpha^2+1072\alpha+960=0;}\\[3mm]
\multicolumn{4}{l}{\hspace*{-1.5mm} \eta(\gamma,\kappa)=
  \frac{2\,\left( 50 - 35\,\gamma  - 8\,{\gamma }^2 + 115\,\kappa  - 
      40\,\gamma \,\kappa  \right) }{15\,\left( -12 + 7\,\gamma  \right) };
   \quad (\gamma_i,\kappa_i) 
  \text{ are the 4 solutions of the following system } 
  (\gamma_i \neq \frac{12}{7})\!:}\\[3mm]
\multicolumn{4}{l}{\hspace*{-1.5mm}\left\{ \begin{array}{l} 
84400 - 12910\,\gamma  - 59329\,{\gamma }^2 + 19128\,{\gamma }^3 + 
  103300\,\kappa  - 116995\,\gamma \,\kappa  + 
  47920\,{\gamma }^2\,\kappa  - 74750\,{\kappa }^2 + 
  26000\,\gamma \,{\kappa }^2=0 \\[3mm]
200 - 380\,\gamma  + 108\,{\gamma }^2 + 530\,\kappa  - 
  64\,\gamma \,\kappa  - 325\,{\kappa }^2 =0 \end{array} \right. } 
\end{array} $$
\end{minipage}}
\closein9 \closein10
\newpage
\thispagestyle{empty}
\enlargethispage{2cm}
\mbox{~~}\\[17mm]
\begin{turn}{90}\usebox{\tabella}\end{turn}
\newpage
\thispagestyle{empty}
\vspace*{10mm}
\begin{turn}{90}\usebox{\tabe}\end{turn}



\begin{thebibliography}{10}
\bibitem{affleck}
I.~Affleck,
J.\ Phys.\ A {\bf 33} (2000) 6473
[arXiv:cond-mat/0005286].
\bibitem{FPR2} 
G.~Feverati, P.~A.~Pearce and F.~Ravanini,
Nucl.\ Phys.\ B {\bf 675} (2003) 469
[arXiv:hep-th/0308075].
\bibitem{F} 
G.~Feverati,
JSTAT {\bf 03} (2004) P001
[arXiv:hep-th/0312201].
\bibitem{LeClair:1995uf}
A.~LeClair, G.~Mussardo, H.~Saleur and S.~Skorik,
theories,''
Nucl.\ Phys.\ B {\bf 453}, 581 (1995) [arXiv:hep-th/9503227].
\bibitem{Dorey:1997yg}
P.~Dorey, A.~Pocklington, R.~Tateo and G.~Watts,
Nucl.\ Phys.\ B {\bf 525}, 641 (1998) [arXiv:hep-th/9712197].
\bibitem{ABF}
G.~E.~Andrews, R.~J.~Baxter and P.~J.~Forrester,
J.\ Stat.\ Phys.\  {\bf 35} (1984) 193.
\bibitem{OPW}
D.~L.~O'Brien, P.~A.~Pearce and S.~O.~Warnaar, Nucl. Phys. B
\textbf{501} (1997) 773.
\bibitem{melzer}
E.~Melzer,
Int.\ J.\ Mod.\ Phys.\ A {\bf 9} (1994) 5753
[arXiv:hep-th/9311058].
\bibitem{BP}
R.~E.~Behrend and P.~A.~Pearce,
J.\ Stat.\ Phys.\ \textbf{102} (2001) 577 
[arXiv:hep-th/0006094].
\bibitem{PearceN}
P.~A.~Pearce and B.~Nienhuis,
Nucl.\ Phys.\ B {\bf 519} (1998) 579
[arXiv:hep-th/9711185].
\bibitem{SY}
R.~Sasaki and I.~Yamanaka,
Adv.\ Stud.\ Pure Math.\  {\bf 16} (1988) 271.
\bibitem{Fioravanti:2002sq}
D.~Fioravanti and M.~Rossi,
JHEP {\bf 0307} (2003) 031
[arXiv:hep-th/0211094].
\bibitem{FinChar}
E.~Melzer,
Int.\ J.\ Mod.\ Phys.\ A {\bf 9} (1994) 1115
[arXiv:hep-th/9305114]; 
A.~Berkovich,
Nucl.\ Phys.\ B {\bf 431} (1994) 315
[arXiv:hep-th/9403073].
\bibitem{zam}
A.~B.~Zamolodchikov,
Adv.\ Stud.\ Pure Math.\  {\bf 19} (1989) 641.
\bibitem{Fioravanti:2003kx}
D.~Fioravanti and M.~Rossi,
JHEP {\bf 0308} (2003) 042
[arXiv:hep-th/0302220].
\bibitem{blz94}
V.~V.~Bazhanov, S.~L.~Lukyanov and A.~B.~Zamolodchikov,
Commun.\ Math.\ Phys.\  {\bf 177} (1996) 381
[arXiv:hep-th/9412229]; Commun.\ Math.\ Phys.\  {\bf 190} (1997) 247
[arXiv:hep-th/9604044]; Commun.\ Math.\ Phys.\  {\bf 200} (1999) 297
[arXiv:hep-th/9805008].
\bibitem{blzxx} 
V.~V.~Bazhanov, S.~L.~Lukyanov and A.~B.~Zamolodchikov,
Nucl.\ Phys.\ B {\bf 489} (1997) 487
[arXiv:hep-th/9607099].
\bibitem{rava}
D.~Fioravanti, F.~Ravanini and M.~Stanishkov,
Phys.\ Lett.\ B {\bf 367} (1996) 113
[arXiv:hep-th/9510047].
\bibitem{kp}
A.~Kl\"umper and P.~A.~Pearce, Physica\ A\ {\bf 183} (1992) 304.
\bibitem{zamoflux} 
A.~B.~Zamolodchikov,
Nucl.\ Phys.\ B {\bf 358} (1991) 524.
\bibitem{Koo:1993wz}
W.~M.~Koo and H.~Saleur,
Nucl.\ Phys.\ B {\bf 426} (1994) 459
[arXiv:hep-th/9312156].
\bibitem{Itoyama:1986ad}
H.~Itoyama and H.~B.~Thacker,
Phys.\ Rev.\ Lett.\  {\bf 58} (1987) 1395.

\end{thebibliography}
\end{document}